\centerline { \bf On Wilson Loops and Confinement without Supersymmetry }
\centerline { \bf from Composite Antisymmeric Tensor Field theories }
\bigskip
\centerline{ Carlos Castro }
\centerline{ Center for Theoretical Studies of Physical Systems }
\centerline { Clark Atlanta University, Atlanta, GA. 30314 }
\centerline { April 2002 }

\bigskip

\centerline{\bf Abstract }

\bigskip

A novel approach that does $not$ rely on supersymmetry , nor in the AdS/CFT
correspondence, to evaluate the Wilson loops asociated with a gauge theory
of area-preserving diffeomorphisms in terms of pure string degrees of
freedoms is presented. It is based on the Guendelman-Nissimov-Pacheva
formulation of composite antisymmetric tensor field theories of
volume-preserving diffeomorphisms. Such theories admit $p$-brane solutions.
The quantum effects are discussed and we evaluate exactly the vev of the
Wilson loops, in the large $N$ limit of  quenched-reduced $SU(N)$ QCD , in
terms of a path integral involving pure string degrees of freedom. It is
consistent with the recent results based on the AdS/CFT correspondence and
dual QCD models ( dual Higgs model with dual Dirac strings ). More general
Loop wave equations in C-spaces ( Clifford manifolds ) are proposed  in
terms of holographic variables that contain  the dynamics of an aggregate of
closed branes ( $ p$-loops ) of various dimensionalities.

\bigskip
\centerline{ \bf I. Introduction }
\bigskip

It has been believed for a long time that QCD confinement is supposed to be
a non-perturbative solution to QCD in four dimensions ; i.e to $SU(3)$
Yang-Mills theory [ 1 ] . A formal proof of the colour confinement amounts
to a derivation of the area-law for a Wilson loop associated, for example,
with the world lines of a quark-antiquark pair joined in by a string . The
area law in the Euclidean regime is
$ W(C) \sim exp [ - T A ] $  where
$ T $ is the string tension and the ( Euclidean ) area is $ A = l t_E $ .
The colour-electric potential rises linearly with the length of the string
separating the quark-antiquark and blows up in the $ l \rightarrow \infty$ .
This would be a signal of ( colour-electric field lines ) confinement, an
infinite amount energy would be required to separate the quarks.
Many attempts have been explored to solve this problem, in particular those
based on the so-called string ansatz [ 2-3 ] :
$$ W[ C ] \sim \int _{\Sigma ( C ) } [ DX ] ~  exp~ ( i S_{string} ) . $$
which says that the effective ( collective ) infrared degrees of QCD at
strong coupling are given by string configurations whose worldsheets have
for boundary the loop $ C$ . The Schwinger-Dyson equations for QCD can be
reformulated as an infinite chain of equations for the Wilson loops that
simplify $drastically$ in the large $ N$ limit giving the single equation
known as the Makeenko-Migdal loop equation [ 4 ] .
In light of the Maldacena AdS/CFT correspondence formulated by many authors
[5 ] as a relation between partition functions, Maldacena and others
proposed that the average value of a Wilson loop in the large $N$ limit, for
$ {\cal N } = 4  ~ SU(N)$ SYM was given by the partition function of a
world-sheet string action  which ends along the loop $ C$ in the
four-dim boundary.
Another approach has been based on the dual formulation of QCD [ 6 ] ( in
the infrared limit ) given by a $ U(1)$ gauge theory adjoined by a dual
Higgs model with dual Dirac strings [ 7  ] ( where the quarks live at their
end-points ) . The average value of the Wilson loop in this dual
phase obeys the area-law fall-off. For other approaches to solve
the confinement problem based on Skyrmions and others methods
see [ 8  ] .

In this work we will present a novel approach that does $not$ rely on
supersymmetry nor the AdS/CFT correspondence, to evaluate the Wilson loop
asociated with a gauge theory of area-preserving diffeomorphisms, in terms
of
the (area ) string degress of freedom. It is based on the
Guendelman-Nissimov-Pacheva formulation of composite antisymmetric tensor
field theories of volume-preserving diffeomorphims [ 9 ] . Such theories
admit $p$-brane solutions after a dualization procedure [ 10 ] .
Our first results are exact on-shell. The quantum effects are discussed next
and
we evaluate exactly the vev of the Wilson loops in the large $N$ limit of
$SU(N)$ QCD in terms of a path integral invoving pure string degrees of
freedom. To achieve this goal we borrow extensively from our results based
on the relationship among large $N$ quenched-reduced $SU(N)$ YM theories and
strings/branes via the Moyal-Zariski deformation quantization [ 11, 16 ] .
This average is consistent with the recent results based on the AdS/CFT
correspondence and dual QCD models ( dual Higgs model with dual Dirac
strings ). Finally we present more general Loop wave equations in C-spaces (
Clifford manifolds )  [ 12, 20  ] than those considered so far.  These loop
equations are given in terms of the holographic variables associated with an
aggregate of closed branes ( $p$-loops ) of various dimensionalities.
\bigskip
\centerline { \bf II }

\centerline {\bf 2.1 Branes as composite antisymmetric tensor field theories
}
\bigskip

In this section we will review the construction of $p'$-brane solutions to
the rank $p+1$ composite antisymmetric tensor field theories [ 10 ]
developed by Guendelman, Nissimov and Pacheva [9 ] when the condition $ D =
p+p' +2 $ is satisfied. These field theories posess an infinite-dimensional
group of volume-preserving diffeomorphisms of the target space of the scalar
primitive field constituents. The role of local gauge symmetry is traded
over to an infinite-dimensional $global$ Noether symmetry of
volume-preserving diffs. The study of the Ward identities for this
infinite-dim global Noether symmetry to obtain non-perturbative information
in the mini-QED models ( the composite form of QED ) was analysed in [ 9 ] .

The starting Lagrangian is defined [10 ] :
$$ L = - { 1\over g^2} F^2_{\mu_1\mu_2...\mu_{p+1}} . ~~~
F = dA = \epsilon_{a_1 a_2....a_{p+1}} \partial_{\mu_1} \phi^{ a_1}
.........
\partial_{\mu_{p+1} }\phi^{ a_{p+1}} . \eqno ( 1 ) $$
the rank $p+1$ composite field strength is given in terms of
$p+1$ scalar fields $ \phi^1(x ) , \phi^2(x) ....\phi^{p+1} (x)$ . Notice
that the dimensionality of spacetime where the field theory is defined is
$greater$ than the number of primitive scalars $ D >  p+1 $.
An Euler variation w.r.t the $\phi^a$ fields yields the following field
equations, after pre-multiplying by a factor of $ \partial_{\mu_{p+2}}
\phi^{ a_1 } $ and using the Bianchi identity $ d F = 0 $ :

$$ \partial_{\mu_1} [ { \delta L \over \delta ( \partial_{\mu_1} \phi^{p+2}
) } ] = 0 \Rightarrow F_{\mu_{p+2} \mu_2...\mu_{p+1}} \partial_{\mu_1}
F^{\mu_1\mu_2...\mu_{p+1}} =0 . \eqno ( 2 ) $$

Despite the Abelian-looking form $ F = dA $ the infinite-dimensional (global
) symmetry of volume-preserving diffs is $not$ Abelian. The theory we are
describing is $not$ the standard YM type .
We are going to find now $p'$-brane solutions to eq-( 2 ) , where $ D =
p+p'+ 2 $. These brane solutions obeyed the classical analogs of $ S$ and $
T$-duality [ 10 ] . Ordinary EM duality for branes requires $ D = p + p' + 4
$. The
latter condition is more closely related to the EM duality among two
Chern-Simons $p,
p'$-branes which are embeddings of a $p, p'$-dimensional object into $ p+2 ;
p'+2 $ dimensions. These co-dimension two objects are nothing but
high-dimensional Knots . For the mathematical intricacies of
Chern-Simons branes, high-dimensional knots and algebraic $K, L $ theory see
[ 13  ] .
A special class of ( non-Maxwellian ) extended- solutions to eqs-( 2 )
requires
a $dualization$ procedure [ 10 ] :
$$ G = * F \Rightarrow G^{\nu_1\nu_2...\nu_{p'+1}} ( {\tilde \phi} (x) ) =
\epsilon^{\mu_1 \mu_2....\mu_{p+1} \nu_1\nu_2....\nu_{p'+1} }
F_{\mu_1\mu_2....\mu_{p+1}} ( \phi ( x ) ) \eqno ( 3 ) $$
After this dualization procedure the eqs- (2) are recast in the form :
$$ G^{\mu_1\nu_2...\nu_{p'+1}}
\partial_{\mu_1} G_{\nu_2\nu_3 ....\nu_{p'+2}} ( {\tilde \phi}  ( x ) ) = 0
. \eqno (
4 ) $$
The dualized equations (4) have a different form than eqs-(2 ) due to the
position of the indices ( the index contraction differs in both cases ).
Extended $p'$-brane solutions to eqs-( 4 ) exist based on solutions to the
Aurilia-Smailagic-Spallucci local gauge field theory reformulation of
extended objects given in [ 14 ] . These are [ 10 ] :
$$ G^{\nu_1\nu_2...\nu_{p'+1}} ( { \tilde \phi} (x) )|_{x = X } = T { \{
X^{\nu_1}, X^{\nu_2} , ......,
X^{ \nu_{p'+1} } \} \over
\sqrt { - { 1 \over (p'+1)! } [ \{ X^{\mu_1} , X^{\mu_2}, ......, X
^{\mu_{p'+1}} \} ]
[ \{ X_{\mu_1} , X_{\mu_2}, ......, X _{\mu_{p'+1}} \} ] } } . \eqno ( 5 )
$$
where $ T$ is the $p'$-brane tension and the Nambu-Poisson bracket w.r.t the
$p'+1$ world-volume variables is defined as the ordinary determinant
/Jacobian :
$$ \{ X^{\nu_1}, X^{\nu_2} , X^{\nu_3} , ......, X^{ \nu_{p'+1}} \}_{NPB} =
\epsilon^{\sigma^1 \sigma^2 \sigma^3....\sigma^{p'+1}} \partial_{\sigma^1}
X^{\nu_1}
\partial_{\sigma^2} X^{\nu_2} .......
\partial_{\sigma^{p'+1}} X^{\nu_{p'+1} } . \eqno ( 6 ) $$
All quantities are evaluated on the world-volume support of the
$p'+1$-brane; i.e. one must restrict the field theory solutions to those
points in the $D$-dimensional spacetime given by $ x = X ( \sigma^1,
\sigma^2,...)$. Solutions to all of the $D$-dim spacetime region can be
extended simply by using delta functionals : $ \delta ( x - X ( \sigma ) ) $
.
\bigskip

\centerline{\bf 2.2 Wilson Loops and Confinement }

\bigskip

In this section we are going to study the string solutions ( $ p' = 1$ ) to
the rank two ( $ p+1 = 2 $ ) composite antisymmetric tensor field theories
of area-preserving diffs in $ D = 4 = p+ p' +2 = 2 + 2 $.
The Wilson loop associated with the composite gauge field is defined :
$$ exp~ [ i \oint_C A_\mu ( \phi^a ) ~ dx^\mu ] .~~~ A_\mu ( \phi )
\equiv \epsilon_{ab} \phi^a (x) \partial_\mu \phi^b ( x ) . \eqno ( 7 ) $$
Due to the Abelian-loooking form of the composite field strength ( as we
said earlier, the algebra of volume-preserving diffs is $not$ abelian ) one
$can$ nevertheless use Stokes law ! :
$$F = dA \Rightarrow F_{\mu\nu} ( \phi ) \equiv \{ \phi^1, \phi^2 \} =
\epsilon_{ab } \partial_\mu \phi^a \partial_\nu \phi^b . ~~~ a , b = 1, 2.
\eqno(8) $$
after using Stokes law the exponential can be written as :
$$ exp ~[ i \int \int _{\Sigma ( C ) } F_{\mu \nu } ( \phi^a ) dx^\mu \wedge
dx^\nu ] . \eqno ( 9 ) $$
where the flux is evaluated through a surface $ \Sigma ( C) $ whose boundary
is $ C$.
if one evaluates all these quantities along the points $x$ whose $support$
lie on the string-world sheet
$x=X$ one may use the string solutions above to the composite antisymmetric
tensor field theory given by the previous equations ( 5 )  :
$$ G ( {\tilde \phi} ) = \Pi = ^* F ( \phi ) \Rightarrow $$
$$ G^{\nu_1 \nu_2 } ( {\tilde \phi} ) |_{ x = X } = \Pi^{\nu_1 \nu_2 } ( X)
= { T \{X^{\nu_1}, X^{\nu_2} \} \over
\sqrt { - { 1 \over 2 } \{ X^\mu , X^\nu \} \{ X_\mu, X_\nu \} } } =
\epsilon^{\nu_1 \nu_2 \mu_1 \mu_2 } F_{\mu_1 \mu_2 } ( \phi )|_{ x = X} .
\eqno ( 10 ) $$
where $ T$ is the string's tension and one is using now ordinary Poisson
brackets. The quantity
$ \Pi^{\mu\nu} $ is the area-conjugate momentum of the string obeying the
Hamilton-Jacobi equation for the string
analog of a point particle momentum.  Hamilton-Jacobi equations for strings
and branes have been given in [ 14 ] . Using these relations above (10)
allows
one to rewrite the flux ( after inserting the product of two spacetime
epsilon tensors $\epsilon_{\mu_1 \mu_2 \mu_3 \mu_4} $ ) as :
$$ { 1 \over 4 ! } \epsilon^{\mu_1 \mu_2 \mu_3 \mu_4 } F_{\mu_1 \mu_2 } (
\phi )
\epsilon_{\mu_1 \mu_2 \mu_3 \mu_4 } d x^{\mu_1 } \wedge dx^{\mu_2 } =
G^{\mu_3 \mu_4 } ( {\tilde \phi} ) d { \tilde \Sigma}_{ \mu_3 \mu_4 } .
\eqno ( 11) $$
For those self dual string configurations , the following relations among
the Poisson brackets are obeyed :
$$ Self~Dual~ Strings \Rightarrow d \Sigma = ^*d \Sigma \Rightarrow
\{ X_{\mu_3} , X_{\mu_4} \}_{PB} = \epsilon_{\mu_1 \mu_2, \mu_3, \mu_4 }
\{ X^{\mu_1} , X^{\mu_2} \}_{PB} . \eqno ( 12) $$
Self dual strings automatically obey the string equations of motion as a
result of the Jacobi identities for the Poisson brackets :
$$ \{ X^\nu , \{ X_\mu , X_\nu \} \} = \epsilon_{\mu\nu \rho\tau} \{ X^\nu ,
\{ X^\rho , X^\tau \} \} = 0 .
\eqno ( 13 ) $$
The vanishing of the second term of the last equation is due to the Jacobi
identities of the Poisson bracket.
Upon evaluation of the flux through the (self-dual) string world sheet ,
whose boundary is $C$ , and restricting to self dual string configurations
allows finally to yield the explicit relationship between the Wilson loop
for the field $ A_\mu ( \phi ) $ and the Dirac-Nambu-Goto string action, in
terms of the string coordinates $ X^\mu (\sigma, \tau ) $, and whose
worldsheet boundary is $ C $ :
$$ W( C ) =exp ~[ i \oint_C A_\mu ( \phi ) dx^\mu ]|_{ x = X } =
exp ~ [ i T \int \int_{\Sigma ( C )} d\sigma d \tau ~
\sqrt { - \{ X^\mu , X^\nu \} \{ X_\mu, X_\nu \} } ] . \eqno ( 14 ) $$
since the determinant of the induced worldsheet metric as a result of the
string's embedding onto the ( flat ) target spacetime is :
$$ det ~ [h_{ab}]  = det~[ \eta_{\mu\nu} \partial_a X^\mu \partial_b X^\nu ]
=
\{ X^\mu , X^\nu \} \{ X_\mu, X_\nu \} . \eqno ( 15) $$
Therefore, we have proven, $on-shell$ , that the Wilson loop associated with
the composite antisymmetric tensor field theory of area-preserving
diffeomorphisms , after using Stokes law, equals the exponential of the
self-dual string action whose worldsheet boundary is $ C$.

Some important remarks are in order. Firstly, one must not confuse the
physical closed string, a loop, with the rectangular Wilson loop associated
with the static quark-antiquark world lines at the ends of an open string of
length $ l $. It is not difficult to verify that the rectangular ( Euclidean
) Wilson loop spanned by the quark-antiquark world lines in a ( Euclidean)
time $ t_E $, the string of length $l$ and of area $ l t_E $, does not obey
the self-dual string equations of motion. Secondly , the physical loop $C$
coincides with the boundary of a closed-string world sheet associated with a
closed string history. For the confinement of $3$ quarks located inside a
$closed$ string using these composite models ( composite measures ) based on
a dynamical tension generation for strings and branes see [ 15 ] .

Notice the importance of the self-dual string condition and that $no$
supersymmetry was required was required nor invoking the AdS/CFT
correspondence was necessary . The only remnants of the AdS/CFT duality were
that the l.h.s of eq- ( 14 ) involves integrating over the closed-string
worldsheet boundary terms, whereas the r.h.s of eq-(14) involves a pure
stringy bulk term ( closed strings contain gravity ) and that a
$dualization$ procedure was necessary in order to find $ p'$-brane solutions
to composite antisymmetric tensor field theories of rank $ p+1$. The fact
that no supersymmetry and that no AdS/CFT was necessay to obtain the
on-shell relation for the Wilson loop in terms of self-dual string
configurations is the first important result of this work.

This has been so far a $classical$ ( on-shell ) result. It is warranted to
study the full quantum theory ( off-shell extension ). In particular to
evaluate the $average$ Wlson loop $ W(C) $ and show, in fact, that it can be
expressed in string variables. This we shall do next.
\bigskip

\centerline{ \bf 2.3  The Average Wilson loop via  Quenched-Reduced  large N
  QCD }
\bigskip
Hoppe long ago [ 22 ] proved that the large $N$ limit ( which is basis
dependent, not unique ) of $ SU(N)$ is isomorphic to the algebra of
area-preserving diffs of a sphere. The topology of the surface is important
since other algebras like $ w_\infty $ and $w_{1 + \infty} $ associated
with $ w_\infty$ strings,  higher conformal spins, are the area-preserving
diffs of a plane and cylinder respectively.  Since the GNP formulation of
antisymmetric tensor field theories involve area-preserving ( volume )
diffs, it is natural to study now the $ SU(N)$ YM theories in the large $N$
limit.

In this last section we will compute the vacuum expectation value of the
Wilson loop in the infinite colour limit  via a Moyal deformation
quantization procedure.  We have shown in [  11 ] that a Moyal deformation
quantization allows to study the large $N$ limit of $SU(N)$  YM theories. $
SU(N)$ reduced-quenched gauge theories admit hadronic bags and Chern-Simons
( dynamical boundaries ) membranes excitations in the large $N$ limit. This
Moyal deformation approach also  furnishes dynamical membranes ( a QCD
membrane )  when the quenching is performed along a $line$, instead of a
point .

Basically, a Moyal quantization takes the Lie-algebra valued operator ${
\hat A } ( x ) $ into a $c$-number
$ A_\mu ( x ; q, p ) $.  Quenching  to a point , and reduction, brings the
quantity $ A_\mu ( x; q. p ) $ to depend solely on the $ (q, p )$
coordinates that subsequently are identified with the internal string/bag
coordinates  after the gauge field-target spacetime coordinate
correspondence
$ A_\mu ( q, p ) \leftrightarrow X_\mu ( \sigma ) $ is made.  Quenching to a
spatial line yields
$ A_\mu ( t; q, p ) $ which are subsequently identified with the membrane
coordinates
$ X_\mu ( t, \sigma^1, \sigma^2 ) $.  Lie-algebra commutators are mapped via
the Weyl-Wigner-Groenowold-Moyal correspondence to Moyal brackets.   The
classical limit $ \hbar \rightarrow 0 $ is related to the large $ N$ limit
via the identification $ \hbar = { 2 \pi \over N } $.  Moyal brackets
collapse to Poisson-brackets in that limit. The Lie-algebra trace operation
is corresponds to an integration w.r.t the string/bag  internal coordinates.

These results can be extended to more general $p$-brane actions given by
Dolan-Tchrakian ( Skyrme type actions ) starting from Generalized Yang Mills
theories in the large $N$ limit ; i.e branes are  roughly speaking Moyal
deformations of Generalized YM theories  [ 11 ] .  Deformation quantization
beyond
the Moyal procedure exists for more generalized Poisson brackets, the
so-called Nambu-Posson brackets which are the Jacobians described earlier.
Its deformation quantization requires the use of the Zariski star product
[ 16]  . It turns out that all $p$-brane actions, including those for
Chern-Simons $p$-branes and Kalb-Ramond couplings to $p$-branes , can be
obtained via a Zariski deformation quantization of the generalized Matrix
models constructed  in  [ 16 ] .  In particular, it was  shown how
Nambu-Goto strings can be obtained directly from $SU(N)$ Born-Infeld models
in the large $N$ limit  [ 17 ] .

To illustrate the power of these approaches we will show how one can obtain
the celebrated Maldacena relation relating the size of the $AdS_5$ throat $
\rho^4$ to the ' t Hooft coupling $ N g^2_{YM} $ and the Planck scale $
L^4_{Planck} \sim ( \alpha )^2 $ ( the inverse string tension squared ) from
a Moyal deformation approach to quenched-reduced large $N$ QCD.  The bag
constant , $\mu$ ,
of mass dimension , was related to the bag tension as  [ 11 ] :

$$ T_{bag } = \mu^4  \sim { 1 \over a^4 g^2_{YM} }. \eqno ( 16 )  $$
where $ a $ was related to the lattice spacing of the large $N$ quenched,
reduced QCD given by
$ ( 2 \pi / a ) = \Lambda_{QCD } = 200$ Mev.  Based on the known result that
a stack of $N$ coincident $ D3$ branes ( whose world volume is
four-dimensional ) in the large $N$ limit is related to black $ p = 3 $
branes solutions to closed type $ II ~ B$ string theory in $ D = 10 $ , and
whose near-horizon geometry is given by $ AdS_5 \times S^5 $  , one may set
the lattice spacing $ a $ associated with large $N$ quenched, reduced $
SU(N) $ YM in terms of the Planck scale $ L_P $ to be $ a^4 = N L^4_P$. This
merely states that we are setting the hadronic bag scale to be $ a = N^{
1/4} L_P $. Inserting this relationship into the expression for the bag
tension gives :

$$ T_{bag } = \mu^4  \sim { 1 \over a^4 g^2_{YM} } =  { 1 \over N L^4_P
g^2_{YM} } \Rightarrow
\mu^{ - 4 } \sim ( N g^2_{YM} ) L^4_P . \eqno ( 17) $$
which has a similar form as the Maldacena relation if one identifies the
size of the
$ AdS_5$ throat to the bag scale  $ \mu^{ -1 } $  . We believe this is $not$
a mere numerical coincidence but stems from the Moyal-Zariski deformation
quantization of the generalized Matrix models [ 16, 17  ] which furnishes
all of the known $p$-brane actions.  Since strings/branes contain gravity
then it is not surpising to see a connection between large $N$ QCD and
gravity.

The average Wilson loop  is defined :

$$ < W _A [ C ] >_{vev}  = \int [ DA ] ~ W _A( C ) ~  e^{ i S_{YM} [ A ] } .
\eqno ( 18 ) $$

The Wilson loop for $ SU(N)$ YM is:

$$ W [ C ] = { 1\over N } trace~Path~ exp~[ i \oint_C A_\mu dx^\mu ]. \eqno
( 19 )  $$

In the quenched-reduced approxomation, defined at a point,  the Wilson loop
shrinks to zero size  and hence the exponential reduces to unity since the
integral has collapsed to zero.
So  then we get  $ W [ C ] \rightarrow  { 1 \over N } trace ~ 1  =  1 $ .
The quenched-reduced  YM action in the large $ N$ limit becomes the
Eguchi-Schild action for the string
after using the $ A_\mu ( \sigma ) \rightarrow X_\mu ( \sigma ) $
correspondence.  This has been known for some time [ 18 ] .

We have then the following  results :

$$ [ DA] \rightarrow [ DX ] ~~~~ W ( C ) \rightarrow 1.  ~~~ e^{ i S_{YM}
}\rightarrow  e^{ i S_{string } }
\eqno  ( 20 ) $$

Under these  conditions  the quenched-reduced  $ SU(N)$ QCD,  in the large
$N$ limit ,  allows to compute exactly the vev of the Wilson loop purely in
terms of string degrees of freedom  given in terms of the Eguchi-Schild
action for the string,  the square of the Poisson brackets, which is
area-preserving diffs invariant :

$$ < W [ C] >_{vev}  =  \int_{ \Sigma (C) } [ DX ] e^{ i S_{string} }
\equiv  \Psi_o [ C ] . \eqno ( 21 ) $$

The physical meaning of this relation can be envisaged as follows. As we
shrink the Wilson loop to a point  , the subsequent large $N$ limit
procedure amounts to introducing an extra dependence on the phase space
variables $ ( q, p )$ , that later are identified as the string coordinates.
   The $ SU(N)$ fiber sitting at the point $ P$ becomes the area world-sheet
of the string in the large $N$ limit. Hence the Wilson loop which had
initially shrunk to a point re-emerges  as an internal loop living in the
$SU(N)$ fiber that was
sitting at the point $P$.  This is compatible with the area-preserving diffs
invariant nature of the Eguchi-Schild action. Roughly speaking, since areas
are preserved, as we shrink the Wilson loop to a point ( to zero )  it must
re-emerged along the fibers in order to preserve the area.

Hence we have obtained an exact  result consistent with those given in the
literature since ( by definition )  the vacuum wave-functional $ \Psi_o [ C
] $ ,  appearing in the r.h.s, is defined by a path integral over all
world-sheets whose boundary is $ C$. The latter is the quantum amplitude for
a closed string to emerge from the vacuum ( a " point " ) and sweep a
world-sheet whose boundary is $ C$. The topology is given by a disc. A
perturbative evaluation of the path integral requires summing over surfaces
of all genera. For more general actions one must restrict  the measure of
integration modulo the volume of the world-sheet diffs group and the group
of Weyl diffs for Polyakov-Howe-Tucker  type of actions .

Nonperturbative effects are never seen in perturbation theory, like the
contribution of self-dual string configurations ( Euclideanized world-sheet
) .  The latter have a dominant weight in the path integral since they
saturate the lower bound of the ( Euclidean ) action.  Hence it is not too
surprising  that the on-shell value of the Wilson loop for composite
antisymmetric tensor field theories was given by the exponential of the
self-dual string action.

The propagator frome one loop configuration $ C_1 $ to another $ C_2 $ is
given by the path integral :

$$ \Delta ( C_1, C_2 ) = \int^{ C_2}_{ C_1 } [ DX] exp ~ [ i S_{string } ] .
\eqno ( 22 ) $$
where the path integral involves summing over all surfaces $ \Sigma ( C_1,
C_2 )$ bounded by the two loops $ C_1, C_2 $.

In [ 19 ] an explicit expression for the string representation of a quantum
loop $ C$,  in terms of a full
$phase$ space string path integral based on the Eguchi-Schild string
action,  was given . The wavefunctional was  a complicated expression : $
\Psi [ x, A, \sigma^{\mu\nu} ] $,
where $\sigma^{\mu\nu}  ( C) $ were the holographic area proyections of a
loop onto the coordinate planes; $A$ was the Eguchi temporal-area variable
associated with the closed loop and $x$ were the coordinate variables of the
loop boundary. What was left $open$ was to see what type of Loop wave
equations the functional  $ \Psi [ x, A, \sigma^{\mu\nu} ] $ obeyed.

Loop wave equations for strings and $p$-branes were given in [ 20 ].
Dirac-like wave equations were also obtained by a suitable " square-root "
procedure which generalized the Hosotoni string Dirac-like equations  [ 21 ]
. One can extend the loop wave equations [ 20  ] to C-spaces ( Clifford
manifolds )
[12 ] by writing the more general loop equations  for a nested family of
$p$-loops, $ p = 0, 1, 2..$
where the maximum value of $p$ corresponds to a spacetime filling brane $ p
+ 1 = D$   :

$$ \Psi [\Omega ,  x^\mu;  \sigma^{\mu\nu} ;  \sigma^{\mu\nu\rho} , .....]
\eqno ( 23 a ) $$
the Clifford-algebra valued object ( a polyvector )  is given by  (
setting the Planck scale to unity  ) :

$$ X = \Omega I + x_\mu \gamma^\mu + \sigma_{\mu\nu} \gamma^\mu \wedge
\gamma^\nu + ....\eqno ( 23 b ) $$
the loop wave equations is :

$$  { \delta^2 \Psi \over \delta \Omega ^2 }    +  { \delta^2 \Psi \over
\ (delta  x_\mu)^2 }   +
{\delta^2  \Psi \over  ( \delta \sigma_{\mu\nu} )^2  }  + {\delta^2  \Psi
\over ( \delta  \sigma_{\mu\nu\rho} )^2 } + ................... = {\cal E}^2
\Psi .    \eqno (
23c  ) $$

The more fundamental problem is to see if these C-space loop equations have
a direct relation to the Maakenko-Migdal loop equations in the infinite
colour limit of YM. The fact the the loop equations in C-spaces incorporate
automatically the closed $p$-brane/ $p$-loops holographic coordinates is
compatible with the AdS/CFT correspondence.

The loop transform ( the analog of the Fourier transform ) for the
full-fledged $SU(N)$ YM theory, in the large $N$ limit , is  :

$$ \Psi [ C ] = \int [ DA ] ~ [ \int d^2 \sigma exp ~ [ \oint_{ C }
~A_\mu
( x^\mu; \sigma) dx^\mu ]  ] ~ \Psi [ A ]. \eqno ( 24 )  $$
where now the gauge field $ A_\mu $ depends on the $x^\mu$ coordinates as
well , in addition to the string variables $ \sigma^1, \sigma^2 $. The
theory is now an effective $higher$ dimensional $ D = 6$  one.  No quenching
nor reduction has taken place in the more general case. Conversely, the
$inverse$  loop transform will yield $ \Psi [ A ] $ in terms of $ \Psi [ C ]
$.  The main question is to see if the  loop equation obeyed by $ \Psi [ C ]
$ agrees in with those given in [ 2 0 ] .

One can generalized these results to $p$-loops  $ C_p$ ; i.e a closed
$p$-brane enclosing a $ p+1$-dimensional region  $\Sigma_{p+1}  ( C_p ) $.
The $p$-loop transform  is defined  in terms of an integral involving the
rank $p$ antisymmetric tensor field  $A_{p} $ :

$$ \Psi [ C_p ] = \int [ DA_{p}  ] ~ \{  \int d^{ p+1}  \sigma ~
exp ~ [  \oint_{C_p } ~A_{\mu_1\mu_2 .....\mu_{ p}}  ( x^\mu; \sigma^1,
\sigma^2,.....)
d \Sigma^{\mu_1 \mu_2 ....\mu_{p} }   ]  \}  ~ \Psi [ A_{p}  ] . \eqno (
25) $$

The integration w.r.t the internal $p+1$ variables of the closed $p$-brane
history ( a $p+1$ worldvolume ) is the analog of the trace operation.  The
antisymmetric tensor field depends on the $x^\mu$ spacetime coordinates and
on the internal $\sigma$ variables.  It is the generalization of the large
$N$ limit of $ SU(N)$ YM given by the $c$-number field $ A_\mu ( x, \sigma )
$ .  One can follow a similar procedure to evaluate the average of the
generalized Wilson loop ; $ W_{A_p} [ C_p ] $ in the quenched-reduced
approximation and express it as a path integral over a  $p$-brane action
whose boundary is given by
$ C_p$.   More details about this will be given in a future publication.

\bigskip

\centerline{ Acknowledgements }

\bigskip

We wish to thank J. Mahecha for his help and to M . Bowers, J. Boedo for
their hospitality in California where this work was completed.

\bigskip

\centerline{ \bf References }

\bigskip

1-K.Wilson : Phys. Rev {\bf D 10} ( 1974 ) 2445.

2-A. Polyakov : Gauge Fields and Strings . harwood Academic Press 1987

3-S. Ketov : Quantum Nonlinear Sigma Models. Springer Verlag 2000.

4- Y. Makeenko, A. Migdal : Nucl. Phys. {\bf B 188 } ( 1981 ) 269.

5-J. Maldacena  : Adv. Theor. Math. Phys.  {\bf 2 } ( 1997 ) 231 .
hep-th/9711200.

S. Gubser, I. Klebanov, A. Polyakov : Phys. Lett {\bf B 428 } ( 1998 ) 105.
hep-th/9802109.

E. Witten : Adv. Theor. Math. Phys. {\bf 2 } ( 1998 ) 253 . hep-th/9802150.

J. Maldacena : Phys. Rev. Lett {\bf 80 } ( 1998 ) 4859. hep-th/9803002

R. Metsaev, A. Tseytlin  : Nuc. Phys. {\bf B 533 } ( 1998 ) 109.
hep-th/9805028.

N. Drukker. D. Gross, A. Tseytlin : JHEP 0004, 2000; hep-th/0001204.

G. Semenoff, K. Zarembo : " Wilson loops in SYM theory : from weak to strog
coupling " hep-th/0202156

S. Bhattacharya : " A short note on the Wilson loop average and the AdS/CFT
corespondence "

hep-th/0202088.

6- M. Baker, J. Ball, F. Zachariasen : Phys. Rep. {\bf 209 }  ( 3 ) ( 1991 )
73.

7-V. Ivanova, N. Troitskaya : " On the Wilson loop in the dual
representation within the dual Higgs model

with dual Dirac strings " hep-th/0112060 .

8-Y. Cho : " Monopole condensation in $ SU(2)$ QCD " hep-th/0201179.

K. Kondo : " A Confining string derived from QCD " hep-th/0110004 .

D. Antonov : " Confining Membranes and dimensional reduction "
hep-th/0107029.

M. Diamantini, C. Trugenberger : " Confining strings at High Temperature "
hep-th/0203053

9 - E. Guendelman, E. Nissimov, S. Pacheva :  " Volume-preserving diffs
versus local gauge symmetry "

hep-th/9505128.

H. Aratyn, E. Nissimov, S. Pacheva : Phys. Lett {\bf B 255 } ( 1991 ) 359.

10 - C. Castro : Int. Jour. Mod. Phys. {\bf A  13 } ( 8 ) ( 1998) 1263-1292

11- S. Ansoldi, C. Castro, E. Spallucci  :  Phys. Lett {\bf B 504 } ( 2001 )
174 .

Class. Quan. Gravity  {\bf 18} ( 2001) L17-L23.

Class. Quan. Gravity  {\bf 18} ( 2001)  2865 .

12- C. Castro :  Chaos, Solitons and Fractals {\bf 12} ( 2001 ) 1585.
physics/0011040.

M.Pavsic  : " The landscape of Theoretical Physics  : A Global view " (
Kluwer, Dordrecht 2001 ).

W. Pezzaglia :  " Physical applications of a generalized Clifford calculus "
gr-qc/9710027.

13- A. Ranicki : " High dimensional Knots , algebraic surgery in codimension
two " Springer Verlag 1998.

14- A. Aurilia, A. Smailagic , E. Spallucci: Phys. Rev. {\bf D 47 } (
1993 ) 2536.

A. Aurilia, E. Spallucci: Class. Quant. Gravity {\bf 10} ( 1993 ) 1217.

15- E. Guendelman: Class. Quan. Grav. {\bf 17 } ( 2000) 3673.
hep-th/0005041.

E. Guendelman, A. Kaganovich, E. Nissimov, S. Pacheva: "String and brane
models with

spontaneoulsy/dynamically induced tension "hep-th/0203024.

16- G. Dito, M. Flato, D. Sternheimer , L. Takhtajan: "Deformation
quantization of nambu Poisson mechanics " hep-th/9602016.

C. Castro: "On the large N limit, $W_\infty$ strings, Star products ...
" hep-th/0106260.

S. Ansoldi, C. Castro, E. Guendelman, E. Spallucci: In preparation:

17- S. Ansoldi, C. Castro, E. Guendelman, Spallucci: "Nambu Goto
strings from $SU(N)$ Born Infeld

Models"  hep-th/0201018 .

18-E. Floratos, J. IIiopulos, G. Tiktopoulos: Phys. Lett {\bf B 217} (
1989) 223.

D. Fairlie, C. Zachos: Phys. Lett {\bf B 224} ( 1989 ) 101.

I. Bars: Phys. Lett {\bf B 245} ( 1990 ) 351.

19- S. Ansoldi, C. Castro, E. Spallucci: Class. Quan. Grav. {\bf 16} (
1999 ) 1833.

20- C. Castro: Chaos, Solitons and Fractals {\bf 11} ( 2000) 1721-1737.

21-Y. Hosotani: Phys. Rev. Lett {\bf 55} ( 1985 ) 1719.

22-J. Hoppe: "Quantum theory of a relativistic surface" MIT Ph.D Thesis
1982

\bye